\documentclass[twocolumn]{article}
\usepackage[a4paper, total={180mm, 243mm}]{geometry}
\usepackage[T1]{fontenc}
\usepackage{balance}
\usepackage{graphicx}
\usepackage{siunitx}
\usepackage{upgreek}
\usepackage{titling}
\usepackage{titlesec}
\usepackage{subcaption}
\usepackage{amsmath, amssymb}
\usepackage{empheq}
\usepackage{xcolor}

\definecolor{authorcolor}{cmyk}{0.8,0.6,0,0.3}

\newcommand{\reffig}[2]{Fig.~\ref{#1}{\textbf{#2}}}

\DeclareCaptionLabelSeparator{custom}{\textbar}
\DeclareCaptionFormat{custom}{\textsf{\textbf{#1 #2} \small #3}}
\newcommand{\ii}{\mathrm{i}}
\renewcommand{\Re}{\operatorname{Re}}
\renewcommand{\Im}{\operatorname{Im}}

\usepackage[
    backend=biber,
    style=nature,
    url=false,
    isbn=false,
    date=year,
    maxnames=99
]{biblatex}
\title{Phase matching and mode coupling in photonic-crystal ring resonators}

\author{Alexander~E.~Ulanov$^{1}$,
        Lukas Bengel$^{2}$,
        Tobias Herr$^{1,3,*}$
}
\date{%
    \small $^1$Deutsches Elektronen-Synchrotron DESY, Notkestr. 85, 22607 Hamburg, Germany \\
    \small $^2$Department of Mathematics, University of Hamburg UHH, Bundesstraße 55, 20146 Hamburg, Germany\\
    \small $^3$Institute of Experimental Physics, University of Hamburg UHH, Luruper Chaussee 149, 22761 Hamburg, Germany\\
    \small $^{*}$tobias.herr@desy.de
}

\begin{document}
\maketitle

\textbf{
Photonic-crystal ring resonators, in which periodic refractive-index modulations couple counter-propagating modes, have become a widely used platform for nonlinear photonics, offering mode-by-mode dispersion engineering. When several modulation components are superimposed, their relative angular alignment (phase) is an additional degree of freedom whose impact on the nonlinear dynamics has remained unclear. Here we show that the relative phases between the modulation patterns have a decisive impact. Through the interference of the nonlinear processes in forward and backward direction, they add a phase and mode selection rule to the established phase-matching conditions, which can fully suppress nonlinear coupling. They also impose a relative spectral chirp between forward- and backward-propagating waveforms. We illustrate these effects for parametric oscillation and dissipative Kerr solitons. Our results extend the concept of phase matching to resonators with interfering nonlinear pathways and establish the alignment of index gratings as a design dimension for nonlinear microresonators.
}

Dielectric microresonators enable efficient access to nonlinear optical effects such as frequency conversion at low pump power~\cite{kippenberg2004KerrNonlinearityOpticalParametric, delhaye2007OpticalFrequencyComb, levy2010CMOScompatibleMultiplewavelengthOscillator}. The efficiency of these processes is governed by phase matching, i.e.\ constructive interference of the nonlinear polarization contributions to the generated wave. In a resonator, phase matching takes the form of a selection rule for the azimuthal mode numbers $m>0$ of the participating modes and a condition on their resonance frequencies. For four-wave mixing (FWM) between input modes $m_1, m_2$ and output modes $m_3, m_4$, the selection rule reads $\Delta m = m_1 + m_2 - m_3 - m_4 = 0$, and the frequency mismatch $\Delta\omega = \left| \omega_{1} + \omega_{2} - \omega_{3} - \omega_{4} + \Delta_\mathrm{NL} \right| \lesssim \kappa$, which must be small compared to the resonance linewidth $\kappa$; $\omega_m$ are the resonance frequencies, and $\Delta_\mathrm{NL}$ lumps together nonlinear resonance shifts due to self- and cross-phase modulation. 
Tailoring the resonance frequencies, and thereby the phase-matching conditions, therefore provides a powerful means of controlling nonlinear processes in microresonators. Established approaches include dispersion engineering of the waveguide cross-section~\cite{okawachi2011OctavespanningFrequencyComb, brasch2016PhotonicChipbasedOptical}, coupling to higher-order transverse modes~\cite{herr2014ModeSpectrumTemporal,liu2014InvestigationModeCoupling, xue2015ModelockedDarkPulse}, and coupled resonators~\cite{helgason2021DissipativeSolitonsPhotonic, tikan2021EmergentNonlinearPhenomena}. 

A particularly versatile approach is offered by one-dimensional photonic-crystal ring resonators (PhCRs)~\cite{arbabi:2011,lu2014SelectiveEngineeringCavity, yu2021SpontaneousPulseFormation, black2022OpticalparametricOscillationPhotoniccrystal}, in which a periodic refractive-index modulation along the resonator circumference selectively couples the degenerate forward- (FWD) and backward-propagating (BWD) modes with a given azimuthal mode number $m$. This coupling hybridizes the two modes and splits their resonance, which strongly modifies the dispersion at the addressed mode. 
Superimposing modulations with different periods enables mode-by-mode dispersion shaping~\cite{lucas2023TailoringMicrocombsInversedesigned, moille2023FourierSynthesisDispersion, chuchalin2026AdjointInverseDesign} and hence fine control of nonlinear processes, which has been used for broadband optical parametric oscillation~\cite{lu2022KerrOpticalParametric, black2022OpticalparametricOscillationPhotoniccrystal, stone2024WavelengthaccurateNonlinearConversion} and for controlled frequency-comb and pulse generation~\cite{yu2021SpontaneousPulseFormation, ulanov2024SyntheticReflectionSelfinjectionlocked, ulanov2026BrightSolitonsHybriddispersion}. Implemented as a lithographically defined corrugation of the waveguide sidewall, such multi-component gratings carry an additional degree of freedom, the relative angular alignment (phase) of their periodic patterns. However, the influence of this alignment on the nonlinear dynamics has remained unclear.

\begin{figure}[t]
  \centering
  \includegraphics[width=\columnwidth]{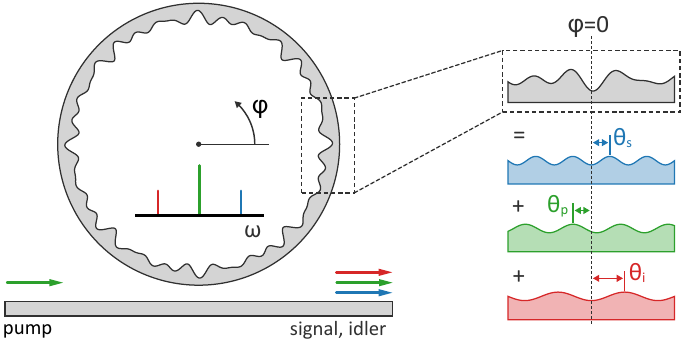}
    \caption{\textbf{Nonlinear frequency conversion in a microresonator with chirped index grating.} A pump laser drives nonlinear frequency conversion inside a microresonators, e.g., the generation of signal and idler sidebands. A periodic corrugation in angular resonator coordinate $\varphi$ of the resonator sidewall creates a periodic refractive index pattern, where multiple corrugation patterns with different spatial period may be superimposed to engineer phase-matching. The index modulation/corrugation patterns may be arranged with different angular phase $\theta$ for pump (p), signal (s), and idler (i) with decisive impact on linear and nonlinear dynamics.
    }
    \label{fig:concept}
\end{figure}

Here, we explore the relative angular alignment of the index modulation patterns (\reffig{fig:concept}{}) as a new degree of freedom to control nonlinear dynamics. We show that their relative phases enter the phase-matching conditions through the interference of the nonlinear processes in forward and backward direction, and that a chirp of these phases, i.e. a mode-dependent angular offset of the components (not of the grating period), imposes a spectral chirp between forward and backward propagating waves. We illustrate this for degenerate FWM and for temporal dissipative Kerr solitons. Our findings establish the alignment of the index modulation patterns as a design dimension for nonlinear microresonators and, more generally, extend the concept of phase matching to systems with multiple interfering nonlinear pathways.

\paragraph{Eigenmodes and linear coupling.} 
In an unperturbed ring, the eigenmodes are the degenerate counterpropagating traveling waves $e^{\ii (\pm m \phi - \omega_m t)}$, where the resonator coordinate (azimuthal angle) $\phi$ is measured from an arbitrary reference point. The intracavity field of azimuthal order $m$ is 
\begin{equation}
E_m(\phi,\tau)=a_m(\tau)\,e^{\ii m\phi}+b_m(\tau)\,e^{-\ii m\phi},
\label{eq:basis}
\end{equation}
where $a_m$ and $b_m$ are the amplitudes of the forward (FWD) and backward (BWD) propagating waves, $\tau=\kappa t/2$ is the normalized time, and $\kappa$ the cavity decay rate. 
A refractive index modulation $\delta n(\phi)$~\cite{lu2014SelectiveEngineeringCavity, lu2022KerrOpticalParametric, black2022OpticalparametricOscillationPhotoniccrystal} with angular period $\pi/m$ satisfies the Bragg condition for the mode $m$ and couples the two degenerate traveling waves $e^{\pm\ii m\phi}$ at a rate $\gamma_m$ set by the modulation depth. Index modulation components with different periods may be superimposed~\cite{lucas2023TailoringMicrocombsInversedesigned,moille2023FourierSynthesisDispersion}, and the generally complex coupling is described by the Fourier transform,
\begin{equation}
\gamma_m=|\gamma_m|\,e^{\ii\theta_m}\propto\oint\delta n_{\mathrm{eff}}(\phi)\,e^{-2\ii m\phi}\,d\phi
,
\label{eq:gamma_def}
\end{equation}
with the phase $\theta_m\equiv\arg\gamma_m$ of the modulation component $\delta n_\mathrm{eff}\propto\cos(2m\phi+\theta_m)$ encoding its angular position relative to the reference point.

The eigenmodes of the coupled system which diagonalize the linear coupling matrix are the hybrid modes
\begin{equation}
\begin{aligned}
E_m(\phi,\tau)=\sum_{\sigma=\pm1}A_m^\sigma(\tau)\,\psi_m^\sigma(\phi),\\
A_m^\sigma(\tau)=\frac{a_m(\tau)-\sigma\,e^{\ii\theta_m}b_m(\tau)}{\sqrt{2}}, \\
\psi_m^\sigma(\phi)=\frac{e^{\ii m\phi}-\sigma\,e^{-\ii\theta_m}e^{-\ii m\phi}}{\sqrt{2}},
\label{eq:hybrid}
\end{aligned}
\end{equation}
with resonance frequencies $\omega_m^{\sigma_m}=\omega_m+\sigma_m|\gamma_m|\,\kappa/2$, with $\sigma_m=+1$ ($-1$) the upper (lower) frequency component. 
The relative phase between BWD and FWD components of each hybrid mode is
\begin{equation}
\label{eq:relphase}
    \arg(b_m/a_m)=-\theta_m+\tfrac{\pi}{2}(1+\sigma_m).
\end{equation}
A differential or chirped phase $\theta_m$ of the index patterns, will therefore give rise to a spectral chirp between forward and backward propagating waves.

\paragraph{Phase matching and nonlinear coupling.}
The coupling between FWD and BWD modes also alters the phase-matching conditions in nonlinear frequency conversion. Here, we consider four-wave mixing (FWM) in a Kerr-nonlinear resonator, but our analysis applies analogously to third-harmonic generation, or the nonlinear processes in resonators with second-order nonlinearity. As a first condition we have 
\begin{equation}
    \label{eq:phcr_pm_1}
    \Delta\omega = \left|\omega_{1}^{\sigma_1} + \omega_{2}^{\sigma_2} - \omega_{3}^{\sigma_3} - \omega_{4}^{\sigma_4} + \Delta_\mathrm{NL} \right| \lesssim \kappa,
\end{equation}
where $\omega_m^{\sigma_m}$ are the hybrid resonance frequencies.

We next compute the mode-overlap integral $\eta \propto \oint\psi_{m_1}^{\sigma_1}\psi_{m_2}^{\sigma_2}\,\psi_{m_3}^{\sigma_3*}\,\psi_{m_4}^{\sigma_4*}\,d\phi$, which is proportional to the nonlinear coupling strength of FWM between the four modes
\begin{equation}
    \label{eq:phcr_pm_2}
    g \propto\delta_{\Delta m,0}\, \left|1+\sigma_1\sigma_2\sigma_3\sigma_4\,e^{-\ii\Delta\theta}\right|
\end{equation}
with $\Delta\theta  =\theta_1+\theta_2-\theta_3-\theta_4$. This 
result may be interpreted as FWM processes that occur independently in FWD and BWD direction, each respecting $\Delta m = 0$, and their interference according to the second factor, which depends on the involved upper and lower frequency modes and $\Delta \theta$. Thus, even if $\Delta m = 0$ and $\Delta \omega = 0$, a complete cancellation of the FWM processes is possible.
In the special case, where FWM occurs within either upper or lower frequency branch (all $\sigma_i$ of the same sign) we obtain
\begin{equation}
    g \propto\big|1+\,e^{-\ii\Delta\theta}\big| = 2\,\big|\cos\!\big(\Delta\theta/2\big)\big| 
\end{equation}

In another special case where all grating components share the same phase (implying a common mirror axis), we obtain
\begin{equation}
    g \propto\delta_{\Delta m,0}\, \left|1+\sigma_1\sigma_2\sigma_3\sigma_4\right|
\end{equation}
which reflects parity conservation in a mirror-symmetric cavity, analogous to the even and odd four-wave-mixing pathways between the supermodes of coupled resonators~\cite{tikan2021EmergentNonlinearPhenomena}.
Below, we provide two examples for illustration.

\begin{figure*}[t]
  \centering
  \includegraphics[width=\textwidth]{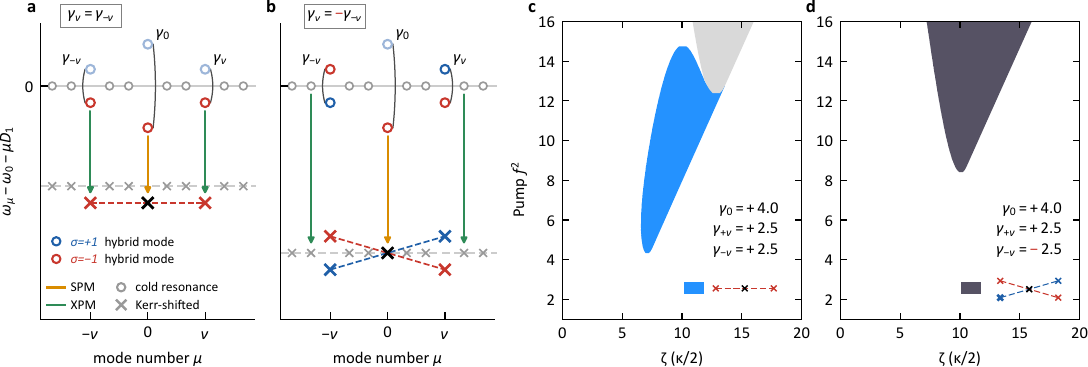}
    \caption{\textbf{Four-wave mixing and modulation instability in a Kerr microresonator with mode hybridization.}
    \textbf{a,b},~Resonance frequencies $\omega_\mu - \omega_0 - \mu D_1$ as a function of mode number $\mu$. Grey circles denote cold-cavity resonances; blue and red circles denote hybrid modes with $\sigma = +1$ and $\sigma = -1$, respectively. In both panels, the lower-frequency ($\sigma = -1$) hybrid resonance of the $\mu = 0$ mode is pumped by a continuous-wave laser. Orange and green arrows indicate the SPM and XPM shifts, and crosses denote Kerr-shifted resonances. 
    \textbf{a},~Equal splitting of the sideband modes, $\gamma_\nu = \gamma_{-\nu}$. The FWD and BWD four-wave-mixing contributions interfere constructively for the sideband pair located on the same side of its doublets as the pump.
    \textbf{b},~Opposite splitting of the sideband modes, $\gamma_\nu = -\gamma_{-\nu}$. Constructive interference occurs for the two degenerate pairs whose sidebands lie on opposite sides of their doublets.
    \textbf{c,d},~Modulation instability (MI) regions as a function of pump detuning $\zeta$ and pump power $f^2$, for pumped-mode splitting $\gamma_0 = 4$ and sideband splitting $\gamma_\nu = \gamma_{-\nu} = 2.5$ (\textbf{c}) and $\gamma_\nu = -\gamma_{-\nu} = 2.5$ (\textbf{d}). Shaded regions mark MI of the type shown in panel \textbf{a} (light blue) and panel \textbf{b} (dark grey). The grey area in panel \textbf{c} highlights a region of oscillation that does arises when the differential nonlinear frequency shifts lift the degeneracy and symmetry between $a_0$ and $b_0$ (cf. main text). All $\gamma_\mu$ are real, $\gamma_0 > 0$, and are given in units of $\kappa/2$.
    }
    \label{fig:opo}
\end{figure*}

\begin{figure*}[t]
    \centering
    \includegraphics[width=\textwidth]{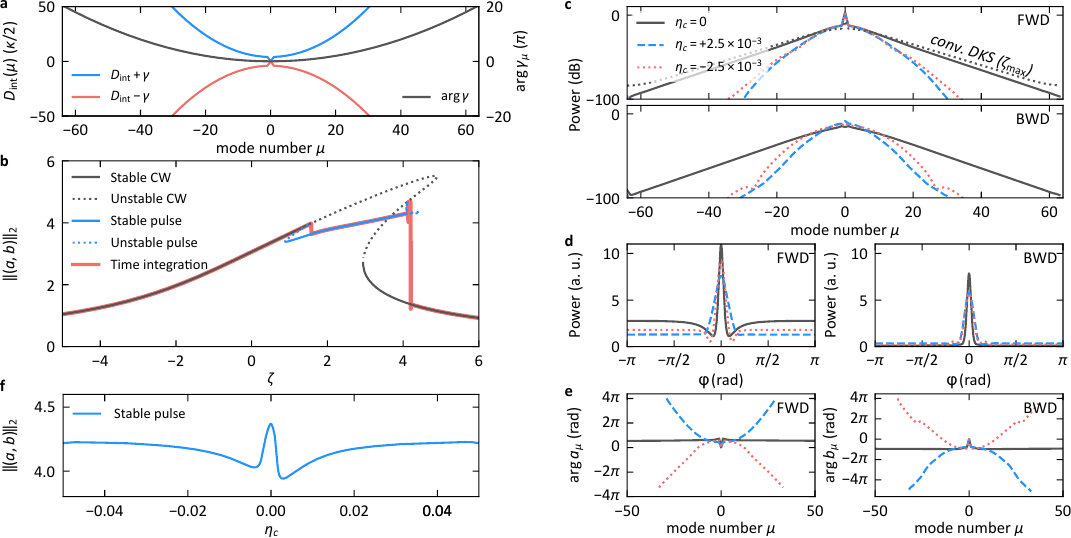}
    \caption{\textbf{DKS generation in a microresonator with chirped index gratings.}
    \textbf{a},~Integrated dispersion of the hybridized modes, $D_\mathrm{int}(\mu) \pm |\gamma_\mu|$ (left axis, in units of $\kappa/2$), and unwrapped grating phase $\arg\gamma_\mu=\theta_\mu$ referenced to $\mu = 0$ (right axis, in units of $\pi$), for $D_\mathrm{int} = 0$, $\gamma_b = 4$, $\gamma_c = 0.05$, and $\eta_c = 2.5\times10^{-3}$.
    \textbf{b},~Norm $\|(a,b)\|_2$ of the stationary states plotted against the normalized detuning $\zeta$ for $f = 2.2$ and the parameters of \textbf{a}, computed by numerical continuation: CW states (black) and pulse states (blue); solid and dotted lines denote stable and unstable solutions, respectively. Red: time integration of Eqs.~\eqref{eq:a}-\eqref{eq:b} with $\zeta$ increased linearly across the pumped resonance.
    \textbf{c},~Optical spectra $|a_\mu|^2$ (FWD, top) and $|b_\mu|^2$ (BWD, bottom) of the pulse states at $\zeta = 2.7$ and $f = 2.2$ for $\eta_c = 0$ (black solid), $\eta_c = +2.5\times10^{-3}$ (blue), and $\eta_c = -2.5\times10^{-3}$ (red); all other parameters as in \textbf{a}. The black dotted line shows a conventional DKS for $f = 2.2$ at $\zeta_\mathrm{max} \approx 6$.
    \textbf{d},~Corresponding intracavity waveforms $|a(\varphi)|^2$ (FWD) and $|b(\varphi)|^2$ (BWD).
    \textbf{e},~Corresponding spectral phases $\arg a_\mu$ (FWD) and $\arg b_\mu$ (BWD).
    \textbf{f},~Norm $\|(a,b)\|_2$ of the pulse state as a function of $\eta_c$ at $\zeta = 2.7$ and $f = 2.2$, obtained by numerical continuation starting from the solution for $\eta_c = 0$.}
    \label{fig:sim}
\end{figure*}

\paragraph{Example 1 -- Optical parametric oscillator.}
As a first example we consider a continuous-wave pumped optical parametric oscillator (OPO) based on degenerate FWM~\cite{kippenberg2004KerrNonlinearityOpticalParametric, matsko2005OpticalHyperparametricOscillations} where a continuous wave pump is 
converted into signal and idler sidebands, which we identify with relative mode indices $\mu$, so that $\mu=0$ is the pumped mode and the signal and idler sidebands are at $\mu =\pm\nu$. 
To visualize the phase-matching condition, we use the integrated dispersion $D_\mathrm{int}(\mu) = \omega_\mu - \omega_0 - \mu D_1$, in which phase-matched mode combinations lie on a straight line; $D_1$ corresponds to the free-spectral range of the resonator. In our case we assume the unperturbed resonator to be dispersion-less, i.e. $D_\mathrm{int}(\mu)=0$, and apply a mode-hybridization to the pumped, signal and idler modes, to isolate the three interacting modes from the other resonator modes; to compensate the differential nonlinear mode-shifts from self- and cross-phase modulation we set $|\gamma_0|>|\gamma_{\pm\nu}|$. This is illustrated in \reffig{fig:opo}{a,b}. The phase matching conditions Eqs.~\ref{eq:phcr_pm_1} and \ref{eq:phcr_pm_2}, now are
\begin{equation}
\label{eq:opo_cond}
\begin{aligned}
        \Delta\omega & = \left|2\omega_{0}^{\sigma_0} - \omega_{\nu}^{\sigma_{\nu}} - \omega_{-\nu}^{\sigma_{-\nu}} + \Delta_\mathrm{NL} \right| \lesssim \kappa,\\
        g & \propto\left|1+\sigma_{\nu}\sigma_{-\nu}\,e^{-\ii\Delta\theta}\right|
\end{aligned}
\end{equation}
with $\Delta\theta=2\theta_0-\theta_{\nu}-\theta_{-\nu}$. We consider two configurations: (i) $\Delta \theta=0$ and (ii) $\Delta \theta=\pi$. Based on Eq.~\eqref{eq:opo_cond} we conclude that in the first configuration, non-zero $g$ is only possible when $\sigma_\nu$ and $\sigma_{-\nu}$ are of the same sign, i.e. both signal and idler sidebands are generated in the same hybrid mode branch ($\{\sigma_\nu, \sigma_{-\nu}\}=\{+1, +1\},\{-1,-1\}$). In contrast, in the second configuration, signal and idler sidebands are generated in different hybrid mode branches ($\{\sigma_\nu, \sigma_{-\nu}\}=\{+1, -1\},\{-1,+1\}$); which of these symmetric oscillation patterns is selected will depend on noise or small asymmetries of the device. 

To confirm these considerations, we perform an analytic stability analysis of the continuous-wave solution, when a CW pump is coupled to the resonator with a certain (normalized) pump power $f^2$ at a certain (normalized) detuning $\zeta$ (see Appendix~\ref{app:stab}), where we assume the pump is only coupled to the FWD $a_0$ mode. As before, we consider two situations, $\Delta \theta=0$ and $\Delta \theta=\pi$, where we set specifically $\gamma_0=4$, $\gamma_{+\nu}=2.5$, and $\gamma_{-\nu}=\pm2.5$. We numerically determine the values $\{f^2,\zeta\}$, where the system shows modulation instability (MI), i.e. the onset of oscillation. To distinguish between the two MI patterns illustrated in \reffig{fig:opo}{a} and \textbf{b}, respectively, we consider two types of unstable eigenvalues: those with $\Re\lambda>0$, $\Im\lambda=0$, and those with $\Re\lambda>0$, $\Im\lambda\neq0$. In the first case, MI produces sidebands on the equidistant frequency grid of the cavity, $\omega_p\pm\nu D_1$. In the second case, the sidebands are offset from this grid and oscillate at $\omega_p\pm\nu D_1\mp\kappa/2\,\Im\lambda$. The resulting stability diagrams, shown in \reffig{fig:opo}{c,d}, confirm the reasoning above. 

For $\Delta\theta=0$ ($\gamma_\nu = \gamma_{-\nu}$; \reffig{fig:opo}{c}), MI occurs within a single hybrid-mode branch, as predicted by the analysis above (light blue). This behavior breaks down at high pump power ($f^2>12$), where differential nonlinear frequency shifts~\cite{delbino2017SymmetryBreakingCounterPropagating} lift the degeneracy of $a_0$ and $b_0$, so that the pump field can no longer be approximated by a single hybrid mode (grey). 
For $\Delta\theta=\pi$ ($\gamma_\nu = -\gamma_{-\nu}$; \reffig{fig:opo}{d}), MI indeed occurs involving both hybrid-mode branches, and the instability region changes substantially (dark grey). First, the threshold power increases. For a mixed-branch pair the nonlinear Kerr-shift $\Delta_\mathrm{NL}$ in Eq.~\eqref{eq:opo_cond} must compensate the full pump shift $|\gamma_0|$, instead of only $|\gamma_0| - |\gamma_\nu|$.
Second, as expected, the two sidebands oscillate in different hybrid branches; we verified numerically that in this case $|\Im\lambda|\approx\gamma_\nu$. These results confirm that the phase relations of the index gratings represented by $\theta_m$ decisively determine the nonlinear dynamics. The next example extends this observation to broadband spectra and ultrashort soliton waveforms.

\paragraph{Example 2 -- Dissipative Kerr solitons.} As a second example, we study the impact of chirped refractive index patterns on dissipative Kerr solitons (DKS) in microresonators~\cite{kippenberg2018DissipativeKerrSolitons, pasquazi2018MicrocombsNovelGeneration, herr2026FrequencyCombsCoherent}. DKS are ultrashort pulsed waveforms that can form under CW driving in a Kerr-nonlinear resonator with anomalous dispersion~\cite{matsko2011ModelockedKerrFrequency,leo2010TemporalCavitySolitons, herr2014TemporalSolitonsOptical}, i.e. where the integrated dispersion is parabolic $D_\mathrm{int}(\mu) = \tfrac{1}{2}D_2\mu^2$ with $D_2>0$. Here, we assume an initially dispersion-free resonator ($D_2=0$), and engineer an anomalously dispersed higher-frequency hybrid mode branch via 
\begin{equation}
    \gamma_\mu = \left(1-\delta_{\mu 0}\right)\left(\gamma_b + \gamma_c\,\mu^2\right)e^{\ii \theta_\mu}.
    \label{eq:gamma_mu}
\end{equation}
where the parameter $\gamma_c$ defines the quadratic dispersion profile of $|\gamma_\mu|$. The factor $\left(1-\delta_{\mu 0}\right)$ leaves the pumped mode $\mu = 0$ unsplit, creating an offset between the anomalous dispersion landscape and the pumped mode, which has been shown to support deterministic single pulse formation~\cite{yu2021SpontaneousPulseFormation, ulanov2024SyntheticReflectionSelfinjectionlocked, ulanov2026BrightSolitonsHybriddispersion}. The phase $\theta_\mu$ does not affect the hybrid resonance frequencies, but it determines the position of the index gratings. For broadband dispersion engineering across many modes, it is an appealing idea to chirp the phase of the index modulation patterns $\theta_\mu = 2\pi \eta_c \mu^2$ with chirp parameter $\eta_c$ to avoid their superposition into a large spike~\cite{lucas2023TailoringMicrocombsInversedesigned}, which would be challenging to fabricate. \reffig{fig:sim}{a} provides an illustration for $\gamma_b = 4$, $\gamma_c = 0.05$, and $\eta_c = 2.5\times10^{-3}$.

To understand the impact of the chirp parameter $\eta_c$, we study the DKS dynamics with two complementary methods: (i) direct time integration of reproduces a pump-frequency scan as observed in experiments see (Appendix~\ref{app:cme}), in which the detuning $\zeta$ is swept slowly across the pumped resonance; (ii) numerical continuation of the stationary solutions, combined with linear stability analysis, yields all stationary stable and unstable states as the detuning is varied. Comparing the two methods identifies which of the stable states are accessed in a pump scan and can therefore be reached directly in an experiment (see Appendix~\ref{app:numerics} for details). 

\reffig{fig:sim}{b} shows the norm $\|(a,b)\|_2 = \big[2\pi\sum_\mu \left(|a_\mu|^2 + |b_\mu|^2\right)\big]^{1/2}$ plotted against the detuning $\zeta$, computed with both methods for a normalized pump amplitude $f = 2.2$ and the dispersion parameters of \reffig{fig:sim}{a}. In the time integration, the intracavity field first follows the upper CW branch until this branch becomes unstable at $\zeta \approx 1.6$, and then transitions directly onto the stable pulse branch~\cite{yu2021SpontaneousPulseFormation, ulanov2024SyntheticReflectionSelfinjectionlocked, ulanov2026BrightSolitonsHybriddispersion}.  
As $\zeta$ increases further, the time-integrated solution coincides with the continued pulse branch. Close to $\zeta \approx 4.1$, where the pulse branch loses stability, the field begins to oscillate and finally decays to the lower CW branch. We repeat these simulations for $\eta_c = 0$ and $\eta_c = -2.5\times10^{-3}$ and, in each case, observe qualitatively similar dynamics and excellent agreement between the two methods. The optical spectra and pulse shapes, however, change significantly with the phase profile of the grating.

\reffig{fig:sim}{c} shows the optical spectra in the forward (top) and backward (bottom) directions at $\zeta = 2.7$ for the three values of $\eta_c$. In each case, the forward and backward spectra have nearly identical, smooth envelopes. For $\eta_c = 0$, the forward spectrum closely matches that of a conventional DKS at the maximum detuning $\zeta_\mathrm{max} = \pi^2f^2/8\approx6$ for $f = 2.2$ (solid and dotted lines, respectively). For $\eta_c = \pm 2.5\times10^{-3}$, in contrast, the spectra are considerably narrower and deviate strongly from the $\mathrm{sech}^2$ envelope characteristic of DKSs in microresonators with anomalous dispersion. The corresponding intracavity waveforms in \reffig{fig:sim}{d} are bright pulses of comparable peak power in both directions, localized at the same angular position. The forward pulse sits on the CW background set by the pump, whereas the backward pulse has only a weak background ($\gamma_0=0$). For $\eta_c \neq 0$, the pulses are broader and have lower peak power than for $\eta_c = 0$; as the system is pumped in the FWD direction (and not symmetrically in both directions), the waveforms for $+\eta_c$ and $-\eta_c$ differ slightly.

The spectral phases $\arg a_\mu$ and $\arg b_\mu$ of these states are shown in \reffig{fig:sim}{e}. For $\eta_c = 0$, the spectral phase is flat, as for a conventional DKS. For $\eta_c \neq 0$, it depends quadratically on $\mu$, with opposite curvature in the forward and backward directions; both curvatures change sign with the sign of $\eta_c$. This follows from Eq.~\eqref{eq:relphase}, which fixes $\arg b_\mu-\arg a_\mu=\pi-\theta_\mu$ for a state in the $\sigma=+1$ branch: the grating chirp must appear as a differential chirp between the two directions, and the pulse cannot be transform-limited in both; this is similar to reflection off of a chirped mirror.
\balance 
Finally, we use $\eta_c$ as the continuation parameter, starting from the solution for $\eta_c = 0$ at $\zeta = 2.7$ and $f = 2.2$, and analyze the linear stability along the resulting branch (\reffig{fig:sim}{f}). The pulse branch remains stable over the entire range $-0.05 \le \eta_c \le 0.05$ and continuously connects the solutions shown in \reffig{fig:sim}{c-e}. 

These results show that the relative phases of the grating components strongly reshape the attractor states of the DKS system. The different spectral chirp between FWD and BWD directions (\reffig{fig:sim}{e}) is a consequence of the linear coupling between FWD and BWD modes, and the sensitivity of the spectra to $\eta_c$ shows that the grating phases decisively shape the nonlinear state; a quantitative account of the soliton dynamics in the presence of the phase-dependent nonlinear coupling is beyond the scope of this work.

\paragraph{Conclusion.}
We have shown that the relative angular phases of the index-modulation components in a photonic-crystal ring resonator, while leaving the resonance spectrum unchanged, enter both the linear coupling between forward and backward waves and the nonlinear coupling between hybridized modes. They impose a relative spectral chirp between forward and backward waveforms, and through the interference of forward and backward four-wave-mixing pathways they add a phase and mode selection rule to the established phase-matching conditions, up to complete cancellation of the nonlinear coupling. We have illustrated this for an optical parametric oscillator, whose threshold and oscillation pattern are set by the grating phases and offer a direct experimental test, and for dissipative Kerr solitons, whose spectra and chirp they reshape. Our findings reveal the importance of the index-modulation phases in Kerr-nonlinear PhCRs and apply analogously to resonators with second-order nonlinearity; they establish the grating phases as a new dimension in the design of nonlinear microresonators.

\bigskip
\appendix
\titleformat{\section}{\normalsize\bfseries}{\thesection}{1em}{}
\titlespacing*{\section}{0pt}{2ex}{1ex}
\noindent\textbf{Appendix}\par\nobreak
\section{System of coupled mode equations}

\label{app:cme}
To simulate nonlinear dynamics, we analyze the coupled mode equations (CMEs)~\cite{chembo2010ModalExpansionApproach} of a dissipative, CW-driven Kerr microresonator in which the FWD amplitudes $a_\mu$ are linearly coupled to the BWD amplitudes $b_\mu$~\cite{skryabin2020HierarchyCoupledMode, kondratiev2020ModulationalInstabilityFrequency}, here with a mode-dependent and complex coupling:
\begin{align}
\begin{split}
\partial_\tau a_\mu &= -(1+\ii\zeta_\mu)a_\mu
 + \ii\!\!\sum_{\mu_1+\mu_2-\mu_3=\mu}\!\! a_{\mu_1}a_{\mu_2}a^{*}_{\mu_3} \\
 &\quad+ 2\ii\Big(\sum_\nu|b_\nu|^2\Big)a_\mu + \ii\gamma_\mu b_\mu + \delta_{\mu0}f,
\end{split}
 \label{eq:a}\\
\begin{split}
\partial_\tau b_\mu &= -(1+\ii\zeta_\mu)b_\mu
 + \ii\!\!\sum_{\mu_1+\mu_2-\mu_3=\mu}\!\! b_{\mu_1}b_{\mu_2}b^{*}_{\mu_3} \\
 &\quad+ 2\ii\Big(\sum_\nu|a_\nu|^2\Big)b_\mu + \ii\gamma^{*}_\mu a_\mu .
\end{split}
 \label{eq:b}
\end{align}
Here the normalized detuning of mode $\mu$ is $\zeta_\mu = 2(\omega_\mu - \omega_p - \mu D_1)/\kappa = \zeta + D_\mathrm{int}(\mu)$, time is normalized as $\tau = \kappa t/2$, the amplitudes are normalized such that $|a_\mu|^2$ and $|b_\mu|^2$ equal the intracavity photon numbers multiplied by $2g_0/\kappa$, where $g_0$ is the Kerr frequency shift per photon, and $f=\sqrt{8g_0\kappa_\mathrm{ex}P_\mathrm{in}/(\kappa^3\hbar\omega_p)}$ is the normalized pump amplitude, with $\kappa_\mathrm{ex}$ the bus-waveguide coupling rate and $P_\mathrm{in}$ the pump power. The factor $2$ in the third terms is the XPM between the counterpropagating fields.

\section{Stability analysis of the three-mode model}
\label{app:stab}
In a three-mode model~\cite{matsko2005OpticalHyperparametricOscillations}, for the pumped resonance $\mu=0$ with all sidebands zero, the stationary form of Eqs.~\eqref{eq:a}--\eqref{eq:b} is
\begin{align}
0&=-\big[1+\ii\delta_a\big]a_0+\ii\gamma_0 b_0+f, & \delta_a&=\zeta-P_a-2P_b,\label{eq:cwa}\\
0&=-\big[1+\ii\delta_b\big]b_0+\ii\gamma_0^{*} a_0, & \delta_b&=\zeta-P_b-2P_a,\label{eq:cwb}
\end{align}
with $P_a=|a_0|^2$ and $P_b=|b_0|^2$. This system determines the stationary FWD and BWD amplitudes, which we write as $a_0 = \sqrt{P_a}\, e^{\ii \alpha}$ and $b_0 = \sqrt{P_b}\, e^{\ii \beta}$. 

Linearizing Eqs.~\eqref{eq:a}--\eqref{eq:b} with respect to the sideband amplitudes $\pm\nu$, and assuming $D_\mathrm{int}(\mu)=0$, yields
\begin{align}
\partial_\tau a_{+\nu}&=-(1+\ii\Delta)a_{+\nu}+\ii\gamma_{+\nu}b_{+\nu}+\ii a_0^2a^{*}_{-\nu},\\
\partial_\tau a_{-\nu}&=-(1+\ii\Delta)a_{-\nu}+\ii\gamma_{-\nu}b_{-\nu}+\ii a_0^2a^{*}_{+\nu},\\
\partial_\tau b_{+\nu}&=-(1+\ii\Delta)b_{+\nu}+\ii\gamma^{*}_{+\nu}a_{+\nu}+\ii b_0^2b^{*}_{-\nu},\\
\partial_\tau b_{-\nu}&=-(1+\ii\Delta)b_{-\nu}+\ii\gamma^{*}_{-\nu}a_{-\nu}+\ii b_0^2b^{*}_{+\nu},
\end{align}
with $\Delta = \zeta - 2(P_a + P_b)$. Conjugating the equations for $a_{-\nu}$ and $b_{-\nu}$ and introducing the state vector $x=(a_{+\nu},\,b_{+\nu},\,a^{*}_{-\nu},\,b^{*}_{-\nu})^{T}$, we obtain $\partial_\tau x=Mx$ with
\begin{equation}
M=\begin{pmatrix}
-(1+\ii\Delta) & \ii\gamma_{+\nu} & \ii a_0^2 & 0\\
\ii\gamma^{*}_{+\nu} & -(1+\ii\Delta) & 0 & \ii b_0^2\\
-\ii a_0^{*2} & 0 & -(1-\ii\Delta) & -\ii\gamma^{*}_{-\nu}\\
0 & -\ii b_0^{*2} & -\ii\gamma_{-\nu} & -(1-\ii\Delta)
\end{pmatrix}.
\label{eq:M}
\end{equation}

Only the relative phase of the two stationary amplitudes matters for the sideband dynamics, therefore we substitute $a_{\pm\nu}\to a_{\pm\nu}e^{\ii\alpha}$ and $b_{\pm\nu}\to b_{\pm\nu}e^{\ii\beta}$. This change of variables does not affect the eigenvalues. With $g=\gamma_{+\nu}e^{\ii(\beta-\alpha)}$, $h=\gamma_{-\nu}e^{\ii(\beta-\alpha)}$, we obtain
\begin{equation}
N=\ii(\tilde M + I)=\begin{pmatrix}
\Delta & -g & -P_a & 0\\
-g^{*} & \Delta & 0 & -P_b\\
P_a & 0 & -\Delta & h^{*}\\
0 & P_b & h & -\Delta
\end{pmatrix},
\label{eq:N}
\end{equation}
where $\tilde M$ is the matrix $M$ written in the new variables and $I$ is the $4\times4$ identity matrix. The eigenvalues $\lambda_M$ of $M$ and $\lambda_N$ of $N$ are related by $\lambda_M=-1-\ii\lambda_N$, i.e.\ $\Re\lambda_M=-1+\Im\lambda_N$, $\Im\lambda_M=-\Re\lambda_N$. The characteristic polynomial of $N$ is
\begin{equation}
\begin{aligned}
\mathcal{P}(\lambda_N) &= \det(N-\lambda_N I)\\
&= \big[(\Delta-\lambda_N)^2-|g|^2\big]\big[(\Delta+\lambda_N)^2-|h|^2\big]\\
&\quad -(\Delta^2-\lambda_N^2)\,(P_a^2+P_b^2)\\
&\quad -P_aP_b\,(gh+g^*h^*)+P_a^2P_b^2 .
\end{aligned}
\label{eq:chi_general}
\end{equation}
It becomes biquadratic in $\lambda_N$ for equal coupling magnitudes, $\gamma_{\pm\nu}=\gamma e^{\ii\theta_{\pm\nu}}$, giving
\begin{equation}
\begin{aligned}
\mathcal{P}(\lambda_N)&=\lambda_N^4+c_1\lambda_N^2+c_0,\qquad S=P_a^2+P_b^2,\\
c_1&=S-2\Delta^2-2\gamma^2,\\
c_0&=(\Delta^2-\gamma^2)^2-S\Delta^2+P_a^2P_b^2-2\gamma^2P_aP_b\cos\Theta,
\end{aligned}
\label{eq:chi}
\end{equation}
with $gh+g^*h^*=2\gamma^2\cos\Theta$ and
\begin{equation}
\Theta=\theta_{\nu}+\theta_{-\nu}+2(\beta-\alpha).
\label{eq:Theta}
\end{equation}
For a pump in a hybrid mode, Eq.~\eqref{eq:relphase} gives $2(\beta-\alpha)=-2\theta_0$ and hence $\Theta=-\Delta\theta$, with $\Delta\theta$ as in Eq.~\eqref{eq:opo_cond}; the grating phases therefore enter the sideband growth rates exactly through the interference phase of Eq.~\eqref{eq:phcr_pm_2}.

The roots of Eq.~\eqref{eq:chi} are $y_\pm=\tfrac12\big(-c_1\pm\sqrt{c_1^2-4c_0}\big)$, and each root gives $\lambda_N=\pm\sqrt{y}$. Writing $\sqrt{y}=u+\ii v$, one obtains $u^2-v^2=\Re y$ and $u^2+v^2=|y|$, and hence $v^2=\tfrac12\big(|y|-\Re y\big)$, $u^2=\tfrac12\big(|y|+\Re y\big)$. With $\lambda_M = -1 - \ii\lambda_N$, the two eigenvalues belonging to one root $y$ are $\lambda_{M,1,2}=-1\pm(v-\ii u)$. Choosing $v\ge0$, the largest growth rate and the associated frequency offset are
\begin{equation}
\begin{aligned}
\max\Re\lambda_M &= -1+\max_\pm\sqrt{\tfrac12\big(|y_\pm|-\Re y_\pm\big)},\\
|\Im\lambda_M|   &= \sqrt{\tfrac12\big(|y_\pm|+\Re y_\pm\big)}
\end{aligned}
\label{eq:lambda}
\end{equation}
where $|\Im\lambda_M|$ is evaluated at the root $y_\pm$ that maximizes $\Re\lambda_M$. Instability requires $\max_\pm\big(|y_\pm|-\Re y_\pm\big)>2$. If $c_1^2\ge4c_0$, both roots are real. If $c_1^2 < 4c_0$, the roots are complex conjugates, $y_-=y_+^*$, and the two unstable eigenvalues form a complex-conjugate pair, $\lambda_{M,1}=\lambda_{M,2}^{*}$. In this case the generated sidebands are displaced from $\omega_p\pm\nu D_1$ by $\pm u\kappa/2$, consistent with energy conservation.

\section{Numerical methods}
\label{app:numerics}

For direct time integration, we truncate Eqs.~\eqref{eq:a}--\eqref{eq:b} to 128 FWD--BWD mode pairs and integrate them with the adaptive third-order Bogacki--Shampine Runge--Kutta scheme (BS3). The detuning $\zeta$ is swept linearly across the pumped resonance at a rate that is low compared with the cavity energy decay rate, so that the intracavity field adiabatically follows the stationary state as long as this state remains stable.

Numerical continuation and bifurcation analysis of the stationary solutions are performed with the path-continuation package pde2path~\cite{uecker2014Pde2pathMatlabPackage}, applied to the time-domain formulation of Eqs.~\eqref{eq:a}--\eqref{eq:b}. The fast time is discretized on a uniform grid of 128 points, and the linear stability of each stationary state is determined from the eigenvalues of the associated Jacobian operator. To ensure robust continuation of localized states, we impose an additional phase condition that removes the continuous translation symmetry in the fast-time variable. For clarity, we show only the DKS branch bifurcating from the CW state and omit all other branches of patterned states. Numerical continuation provides a systematic picture of the phase space of Eqs.~\eqref{eq:a}--\eqref{eq:b}, whereas time integration shows which of the stable states are reached in a pump scan.

\paragraph{Funding}
This project has received funding from the European Research Council (ERC grant agreement No 101293333), the European Innovation Council (EIC grant agreement No 101137000), and the Helmholtz Young Investigators Group VH-NG-1404; the work was supported through the Maxwell computational resources operated at DESY.

\printbibliography

\end{document}